\documentstyle[prb,aps,twocolumn]{revtex}

\begin{document}
\draft \flushbottom

\twocolumn[\hsize\textwidth\columnwidth\hsize\csname
@twocolumnfalse\endcsname

\title{Electronic properties and Fermi surface of Ag(111) films deposited
onto H-passivated Si(111)-(1x1) surfaces}
\author{A. Arranz}
\address{LURE, Centre Universitaire Paris-Sud, Bat. 209 D, B.P. 34, 91898
Orsay Cedex, France\\
and Dpt. F\'{\i}sica Aplicada, Fac. de Ciencias, C-XII, Univ.
Aut\'{o}noma de Madrid, Cantoblanco, 28049 Madrid, Spain}
\author{J.F. S\'{a}nchez-Royo}
\address{LURE, Centre Universitaire Paris-Sud, Bat. 209 D, B.P. 34, 91898
Orsay Cedex, France\\
and Dpt. F\'{\i}sica Aplicada, ICMUV, Univ. de Valencia, c/Dr.
Moliner 50, 46100 Burjassot, Valencia, Spain}
\author{J. Avila}
\address{LURE, Centre Universitaire Paris-Sud, Bat. 209 D, B.P. 34, 91898
Orsay Cedex, France\\
and Instituto de Ciencia de Materiales de Madrid, CSIC,
Cantoblanco, 28049 Madrid, Spain}
\author{V. P\'{e}rez-Dieste and P. Dumas}
\address{LURE, Centre Universitaire Paris-Sud, Bat. 209 D, B.P. 34, 91898
Orsay Cedex, France}
\author{M.C. Asensio\cite{byline}}
\address{LURE, Centre Universitaire Paris-Sud, Bat. 209 D, B.P. 34, 91898
Orsay Cedex, France\\
and Instituto de Ciencia de Materiales de Madrid, CSIC,
Cantoblanco, 28049 Madrid, Spain}

\date{\today}
\maketitle
\begin{abstract}

Silver films were deposited at room temperature onto H-passivated
Si(111) surfaces. Their electronic properties have been analyzed
by angle-resolved photoelectron spectroscopy. Submonolayer films
were semiconducting and the onset of metallization was found at a
Ag coverage of $\sim$0.6 monolayers. Two surface states were
observed at $\overline{\Gamma}$-point in the metallic films, with
binding energies of 0.1 and 0.35 eV. By measurements of
photoelectron angular distribution at the Fermi level in these
films, a cross-sectional cut of the Fermi surface was obtained.
The Fermi vector determined along different symmetry directions
and the photoelectron lifetime of states at the Fermi level are
quite close to those expected for Ag single crystal. In spite of
this concordance, the Fermi surface reflects a sixfold symmetry
rather than the threefold symmetry of Ag single crystal. This
behavior was attributed to the fact that these Ag films are
composed by two domains rotated 60$^o$.

\end{abstract}

 ]

\section{Introduction}

Among other metal-silicon systems, the growth of silver (Ag) onto
Si(111) surfaces is one of the most extensively
studied.\cite{AgSiH1} Typically, Si(111) surfaces exhibit a wide
variety of reconstructions, originated by the tendency to saturate
the dangling bonds resulting from the abrupt surface termination.
Because of this, deposition of Ag adatoms onto these surface
reconstructions results in Ag/Si(111) interfaces with strongly
different electronic properties of the interface. By far, the most
commonly analyzed Ag/Si(111) systems are those of Ag films onto
Si(111)-(7x7) and the Ag-derived
Si(111)-($\sqrt{3}$x$\sqrt{3}$)R30$^{o}$ surface reconstruction.
The interest in such systems lies on the preparation of well
ordered structures which show new applications in optoelectronics
as magnetic or optical devices. In general, it is highly desirable
to achieve sharp interfaces, in which the conductivity and
transport properties depend on their size and spatial
distributions. Nevertheless, because of stress effects, the most
common growth mode at room temperature (RT) for nonreactive
Ag/Si(111) interfaces is the formation of three-dimensional
islands with widely varying
heights.\cite{AgSiH2,AgSiH3,AgSiH4,AgSiH5} In order to avoid these
effects and to improve the conductivity and transport properties
of the Ag films, recent studies have focused their efforts on
alternative growth modes and regimes.\cite{AgSiH6,AgSiH7,AgSiH8}

In this context, much effort has been devoted to find mechanisms
that neutralize dangling bonds at the Si(111)
surface.\cite{AgSiH9,AgSiH10} With this purpose, interest has been
recently revived by the obtention of artificially produced
H-terminated surfaces with a high degree of homogeneity. The
hydrogenation of the Si(111)-(7x7) restores the (1x1) symmetry of
the bulk and seems to destroy the Fermi level (E$_F$) pinning of
the non-hydrogenated surface.\cite{Aminabis,Amina} The development
of both the wet chemical treatment and the atomic hydrogen-based
method has allowed the preparation of a stable and unreconstructed
flat Si(111) surface,\cite{AgSiH11,AgSiH12} with a
nearly-defect-free termination. As a consequence of the
high-quality of this hydrogenated surface, four remarkably sharp
features have been recently resolved by angle-resolved
photoelectron spectroscopy (ARPES), which appear to be mostly
located around the
$\overline{K}$-point.\cite{AgSiH12,AgSiH13,AgSiH14} The origin of
these features is attributed to: (i) A surface resonance, with a
p$_{x}$-p$_{y}$ symmetry,\cite{AgSiH14} (ii) A surface state
identified as a Si-Si backbond state,\cite{AgSiH14} and (iii) two
higher binding-energy H-Si surface states.\cite{AgSiH14,AgSiH15}

In spite of the extensive investigations of the electronic
properties of Ag films deposited onto clean Si(111) substrates,
there remains a lack of studies on the electronic properties of Ag
films deposited onto H/Si(111)-(1x1) substrates. The study of the
Ag films deposited onto these surfaces has been mostly approached
through structural methods,\cite{AgSiH16,Naitoh3,Sakurai} in order
to elucidate their growth mode on these passivated surfaces.
Growth mode and structure of these Ag films is drastically changed
by the hydrogen termination of the Si(111)-(7x7)
surface.\cite{AgSiH16,Naitoh3} It is well-known that Ag films
deposited onto clean Si(111) surfaces grow in quasi-layer-by-layer
fashion up to few monolayers (MLs) at RT, with a two-domains
Ag-islands distribution.\cite{dosdominios} At high substrate
temperatures, growth of Ag films onto clean Si substrates proceeds
in the Stranski-Krastanov mode.\cite{AgSiH2} Opposite to this
behavior, impact-collision ion-scattering spectroscopy showed that
growth of Ag films deposited onto H/Si(111)-(1x1) surfaces follows
a quasi-layer-by-layer mode even at temperatures of 300
$^{o}$C.\cite{AgSiH16} In this case, the forming Ag islands
appeared to be thinner than those deposited onto the
high-temperature clean substrates.\cite{AgSiH16,Naitoh3} Moreover,
it was observed that Ag films deposited onto H/Si(111)-(1x1)
substrates at 300 $^{o}$C tend to follow a mono-domain island
distribution.\cite{AgSiH16}

With regard to the presence of H at the interface, it has been
shown that its presence gradually decreases with the Ag deposition
even at RT,\cite{AgSiH17,AgSiH18} indicating a partial replacement
of H by Ag atoms. This effect has been recently confirmed by
resonant nuclear analysis of the Ag-H/Si(111) interface in 1 ML
passivated Si substrates prepared by dosing atomic H at 820
K.\cite{AgSiH19} These results show that low-temperature
deposition of Ag causes part of the H-monolayer migrates to the
surface and the rest remaining at the interface. In contrast,
deposition at 360 K prevents the presence of H at the metal
surface.

In this work we report on the electronic properties of Ag films
deposited onto H/Si(111)-(1x1) surfaces at RT, determined by
ARPES. In Sect. II we describe the experimental setup. The results
obtained by ARPES are showed and discussed in Sect. III. In this
section we illustrate and discuss the evolution of the Ag
valence-band structure as the film thickness increases and the
influence of H on it. We finish this section with the analysis of
the Fermi surface (FS) of the metallic films prepared.

\section{Experiment}

The experiments were performed at LURE (Orsay, France) using the
French-Spanish (PES2) experimental station of the Super-Aco
storage ring, described elsewhere.\cite{AgSiH20} The measurements
were carried out in a purpose-built ultra-high vacuum system, with
a base pressure of 5x10$^{-11}$ mbar, equipped with an angle
resolving 50 mm hemispherical VSW analyzer coupled on a goniometer
inside the chamber. The manipulator was mounted in a two-axes
goniometer that allows rotation of the sample in: (i) The whole
360$^{o}$ azimuthal angle ($\phi$) and (ii) In the 180$^{o}$ polar
emission angle relative to surface normal ($\theta_{off}$), with
an overall angular resolution of 0.5$^{o}$. The energy range of
light was of 18-150 eV. Photoelectrons were excited with
p-polarized synchrotron radiation, that is, with the polarization
vector of incident light, the surface normal, and the emitted
electrons lying in the same plane. The current incident angle of
the light ($\Theta_{i}$) was $\Theta_{i}$=45$^{o}$. Nevertheless,
some experiments were carried out as a function of $\Theta_{i}$.

With this set-up, the procedure to determine the FS using ARPES is
direct. For a given experiment, the photon energy ($h\upsilon$)
was fixed and the intensity at the E$_F$ was recorded, along a
series of azimuthal scans, for each step of the crystal rotation
about its surface normal. This procedure was repeated at different
polar angle positions of the analyzer, which allows us to scan a
sheet of the Brillouin zone (BZ) for each h$\upsilon $. In order
to measure the FS contour at different perpendicular wave vectors,
the above procedure can be carried out at different photon
energies. In our measurements, the typical polar intervals were
1.5$^{o}$ and the azimuthal angle range was fixed at 180$^{o}$. In
these conditions, the typical measuring time was between 4 and 6
hours, depending on statistics and optimal signal-to-noise ratio.

The substrates were n-doped Si(111) single crystal, with a nominal
resistivity of 100 $\Omega$cm. They were prepared \textsl{ex situ}
using a wet chemical treatment that results in a passivated
H/Si(111)-(1x1) surface.\cite{AgSiH11} After introducing the
substrates in the chamber, the quality of the surfaces was checked
out through the sharpness of the features appearing in the
valence-band photoemission-spectra at $\overline{K}$-point, which
are attributed to intrinsic H-Si surface
states.\cite{AgSiH13,AgSiH14} Ag was evaporated onto the surface
at RT. The rate of evaporation, 0.06 ML/min, was determined by
using a quartz microbalance. In these conditions, Ag films of
thickness ranging from 0.03 to 8 ML were deposited onto
H/Si(111)-(1x1) surfaces at RT.

\section{Results and Discussion}

\subsection{Metallization of Ag films deposited on H/Si(111)-(1x1)
surfaces}

Ag films were deposited at RT onto H/Si(111)-(1x1) surfaces with a
nominal Ag coverage ($\theta_{Ag}$) lower than 1 ML. Valence-band
photoemission measurements were carried out in these films.
Figure~\ref{fig1} shows the energy distribution curves (EDCs)
measured in these films with h$\upsilon$=32 eV at $\overline{K}$
and $\overline{\Gamma}$-points of the Si(111) surface Brillouin
zone (SBZ). All the spectra are referred, in binding energy, to
the E$_F$. With regard to the spectra corresponding to clean
substrate, the most remarkable features at $\overline{K}$-point
are those appearing with binding energies between 4 and 10 eV
(Fig.~\ref{fig1}(a)). These features have been related with
emission from surface states ((a), (a'), and (b`) features) and a
surface resonance ((b)-feature).\cite{AgSiH13,AgSiH14} Opposite to
this, main emission from the substrate at
$\overline{\Gamma}$-point is related to bulk Si transitions.

As the $\theta_{Ag}$ increases, the substrate features at both
$\overline{K}$ and $\overline{\Gamma}$-points tend to be
attenuated by the film signal. This effect is expected to be
stronger for surface states than for bulk-derived states or
surface resonances, due to the intrinsic localization of the
surface states.\cite{AgSiH14} This would explain the faster
decrease of intensity of (a) and (a`) features compared to that of
the (b)-feature with $\theta_{Ag}$ (Fig.~\ref{fig1}(a)), which
would support the assignment of the (b)-feature as a surface
resonance state.\cite{AgSiH14}

In spite of Ag deposition, non-appreciable changes in binding
energy and line width of the substrate-derived features are
detected as the $\theta_{Ag}$ increases. On the other side, no
trace from Ag-induced surface states at these points of the SBZ
were detected. These facts suggest that this metal-semiconductor
interface shows a nearly abrupt and nonreactive behavior.

The insets of Fig.~\ref{fig1} show in detail the evolution with
$\theta_{Ag}$ of the density of states (DOS) at the E$_F$ in these
submonolayer Ag films. At $\overline{\Gamma}$-point
(Fig.~\ref{fig1}(b)), the DOS at the E$_F$ is negligible and the
small emission observed can be attributed to secondary electrons.
The only emission expected to appear close to the E$_F$ at
$\overline{\Gamma}$-point in metallic Ag(111) films would come
from the Shockley Ag(111) surface state. In Ag films deposited
onto clean Si substrates, emission from this state appears in
films thicker than 3 MLs.\cite{AgSiH21} Therefore, the lack of
emission at $\overline{\Gamma}$-point with h$\upsilon$=32 eV is
not really surprising in such nonreactive interfaces. At
$\overline{K}$-point, the situation is completely different from
that observed at $\overline{\Gamma}$-point. This symmetry point
has a parallel wave vector (k$_\|$) of k$_\|$=1.09 \AA$^{-1}$,
which appears to be slightly smaller than the Fermi vector (k$_F$)
of the bulk Ag(111) FS cut with h$\upsilon$=32 eV along the
$\overline{\Gamma}$$\overline{K}$ direction (k$_F$=1.13
\AA$^{-1}$).\cite{AgSiH22} Therefore, the evolution of the DOS at
the E$_F$ at this symmetry point is directly related with the
metallization process in these films due to the \textsl{sp}-band
crossing the E$_F$. The inset of Fig.~\ref{fig1}(a) elucidates
this point. It can be observed that the DOS at the E$_F$ increases
with the $\theta_{Ag}$ with an incipient \textsl{sp}-derived state
close to the E$_F$ at $\theta_{Ag}$=0.56 ML. This can be
considered the onset of metallization of Ag films deposited onto
H/Si(111)-(1x1) surfaces. These results would confirm the
nonreactive behavior of Ag overlayers and suggest that
metallization process of Ag films prepared onto H/Si(111)-(1x1)
surfaces at RT follows the tendency through bulk-like metallic
films in a similar way to that observed in Ag films prepared onto
clean Si surfaces at RT,\cite{AgSiH21} without additional
metal-semiconductor surface states induced at the E$_F$.

As it is expected for a quasi-layer-by-layer growth of Ag films
onto H/Si(111)-(1x1) substrates at RT,\cite{AgSiH16} the whole
metallization process can be monotonously followed in films
thicker than 1 ML. Figure~\ref{fig2} shows, in a similar way, the
EDCs measured in films thicker than 1 ML. The main features
observed in these spectra at both symmetry points correspond to
the Ag \textsl{4d} valence band, appearing at binding energies
higher than 4 eV. The Ag \textsl{4d} features are better resolved
than those measured in Ag films with similar thickness
as-deposited onto clean Si substrates. This fact can be related to
the different growth mode of these films, which is an indicative
of the structural improvement obtained in films prepared onto
H/Si(111)-(1x1) surfaces at RT. The insets of Fig.~\ref{fig2} show
the evolution of the DOS at the E$_F$ with $\theta_{Ag}$. At
$\overline{K}$-point, the gradual increasing of emission at the
E$_F$ observed in the submonolayer films defines in a clear peak.
At $\theta_{Ag}$=2.4 ML, the bulk-like long lifetime
\textsl{sp}-band is formed and, consequently, the metallic
behavior of these Ag films is completely established at this Ag
coverage. At $\overline{\Gamma}$-point, the evolution of the DOS
at the E$_F$ is completely different from that observed in the
submonolayer films. In addition, Ag films of round to 2 ML show a
new incipient emission in the vicinity of the E$_F$ that becomes
resolved in thicker films as two features, labelled as SS1 and
SS2. These features correspond to occupied states SS1 and SS2 with
binding energies of 0.10 eV and 0.35 eV, respectively. The origin
of these two features will be discussed in the next section.

\subsection{Surface states of metallic Ag films}

In the previous section, it was pointed out the existence of SS1
and SS2 features in the valence band spectra at
$\overline{\Gamma}$-point of Ag films deposited at RT onto
H/Si(111)-(1x1) with $\theta_{Ag}>$2 ML. Now, we discuss the
origin of the states that give rise to these features.
Figure~\ref{fig3} shows the normal-emission valence-band spectra
measured with h$\upsilon $=32 eV in films prepared with different
conditions. In this Figure, curves \textsl{a}-\textsl{c}
correspond, respectively, to 4, 5, and 6 ML-thick as-deposited Ag
films. Curve \textsl{d} corresponds to the latter 6 ML-thick Ag
film annealed to 300 $^{o}$C for 5 min. The latter sample was
annealed again to 300 $^{o}$C for 15 min and then 1 ML Ag film was
deposited (Curve \textsl{e}). Curve \textsl{f} corresponds to a
further Ag deposition of 1 ML and then annealed to 300 $^{o}$C for
5 min. The binding energies of the SS1 and SS2 states are marked
by solid lines in these spectra. From these results, we can remark
that: (i) These states appear with Ag deposition onto
H/Si(111)-(1x1) surfaces in films thicker than 2-3 MLs and their
intensities simultaneously increase with $\theta_{Ag}$. (ii)
Annealing effects make emission from the SS2 feature to decrease,
whereas the SS1 feature is enhanced. Moreover, it appears that
this annealing effect is irreversible with additional Ag
deposition. (iii) The binding energy of the SS1 state is nearly
constant in the whole Ag film-thickness range studied.
Nevertheless, the SS2 feature appears slightly shifted to lower
binding energy after annealing.

Figure~\ref{fig4} shows the valence-band spectra measured by ARPES
along the [$\overline{1}$10] direction with h$\upsilon$=32 eV in a
7 ML-thick Ag film deposited at RT onto H/Si(111)-(1x1) and
annealed to 300 $^{o}$C for 20 min. Similar results are obtained
for the spectra measured along the [$\overline{2}$11] symmetry
direction (not shown). Small solid bars indicate the position of
the SS1 and SS2 states as they disperse. The inset shows the band
diagram as extracted from the dispersion of these features. The
SS1 and SS2 states appear to cross the E$_F$ at k$_{\Vert }$=0.16
and 0.22 \AA$^{-1}$, respectively. Solid lines are parabolic
fitting curves of the obtained dispersions, which correspond to an
in-plane effective mass (m$^{*}_{\Vert }$) of m$^{*}_{\Vert
}$=0.88m$_o$ and 0.56m$_o$ for the SS1 and SS2 states,
respectively, being m$_o$ the free electron mass.

Dispersion of SS1 and SS2 states along the [111] symmetry
direction has been also determined. Figure~\ref{fig5} shows the
normal-emission valence-band spectra measured in a 6 ML-thick Ag
film deposited at RT onto H/Si(111)-(1x1) and then annealed to 300
$^{o}$C for 5 min, as a function of the h$\upsilon$. Solid lines
indicate the binding energies of the SS1 and SS2 states in these
spectra. From these results, it can be observed that these states
do not disperse in the perpendicular wave vector range studied,
which extends over the whole $\Gamma$L symmetry direction of the
Ag(111) BZ.

The facts observed are: (i) SS1 and SS2 show a parabolic in-plane
dispersion (Fig.~\ref{fig4}) and (ii) they do not disperse with
h$\upsilon$ (Fig.~\ref{fig5}), indicating clearly that SS1 and SS2
states can be considered as two-dimensional states. The first
state (SS1) can be identified as the well-known Shockley surface
state of bulk Ag(111) (or \textsl{sp}-surface state), whose
existence is due to the break of crystalline periodicity at the
surface.\cite{Echenique} This state is located in the {\sl sp}
band gap at the L-point of the bulk BZ. ARPES measurements carried
out in Ag(111) single crystal show that this state disperses as a
parabolic surface state centered at $\overline{\Gamma}$-point,
with a maximum binding energy of $\sim$0.12 eV and crossing the
E$_F$ at k$_{\Vert }$=0.14 \AA $^{-1}$.\cite{McDougall} On the
other side, This state has been observed in Ag films deposited
onto clean Si(111) and Cu(001), when these films are thicker than
2-3 MLs.\cite{AgSiH21,ss1} All these facts are in concordance with
our observations and support the assignment of SS1 as the Ag(111)
\textsl{sp}-surface state.

With regard to the two-dimensional SS2 state, it appears at 0.25
eV below the SS1 surface state. This state is rather confined in
the surface and spreads along the surface. At first sight, this
state may be tentatively assigned to be a Ag(111)
\textsl{sp}-derived quantum-well state. Shallow quantum states
with m$^{*}_{\Vert }\sim$0.4m$_o$ have been observed at binding
energies of 0.5 eV,\cite{qw,chiangGL} which is, in fact, quite
close the top of the Ag \textsl{sp}-band in the
$\Gamma$L-direction.\cite{chiangGL,bandstruct} Nevertheless, this
assignment of the SS2 state seems not to be correct, since it is
not expected quantum states at binding energies of 0.35 eV in
2ML-Ag films and this state appears slightly outside the
\textsl{sp}-band range in the Ag $\Gamma$L direction. In addition,
its binding energy appears to be constant with $\theta_{Ag}$ in
the thickness range of the Ag films studied. In fact, a shallow
Ag(111) \textsl{sp}-quantum state should show a shift of $\sim$0.2
eV to lower binding energies in the Ag thickness range of 4-8 MLs,
following the phase-accumulation-model.\cite{phaseaccmod}

After these considerations, it appears that the SS2 state is
related to a Ag-derived surface state or surface resonance, since
it appears quite close, in binding energy, to the bottom of the
\textsl{sp}-gap of the Ag valence-band projected onto the
(111)-face. From our results (Fig.~\ref{fig2} and
Fig.~\ref{fig4}), we have observed that SS2 emission increases in
a similar way to that of SS1 as a function of the Ag deposition.
In addition to this, this state appears to be a parabolic state
with a m$^{*}_{\Vert }$ quite close to that obtained for the
\textsl{sp}-surface state of bulk Ag(111).\cite{masaeffsurfstate}
These results suggest that SS2 is a surface state related to a
bulk-like termination of the Ag(111) films, as the SS1 state,
shifted to higher binding energy.

In order to check these assignments for both SS1 and SS2 states,
we have analyzed the behavior of their wave functions as a
function of $\Theta_{i}$. By means of $\Theta_{i}$, the weight of
the components of the polarization vector of light can be tuned,
either along the normal to the surface (large $\Theta_{i}$) or
contained in the surface plane (small $\Theta_{i}$). In this way,
emission from perpendicular-to-surface components of the wave
function is enhanced at large $\Theta_{i}$ and decreased at small
$\Theta_{i}$. In the case that the emission plane is a mirror
plane, p-polarized excitation selects even states with respect to
this plane. As we are interested only in \textsl{sp}-states, even
states refer to \textsl{p}-states lying in the emission plane and
\textsl{s}-states. Therefore, at large $\Theta_{i}$, it would be
expected an enhancement of their \textsl{p$_z$}-character.

Figure~\ref{fig6} shows the normal-emission valence-band spectra
measured with h$\upsilon$=50 eV in a 6 ML-thick Ag film deposited
onto H/Si(111)-(1x1) and then annealed to 300 $^{o}$C for 5 min,
as a function of the $\Theta_{i}$. The emission plane is that
determined by the $\overline{\Gamma}$$\overline{K}$-direction,
which can be considered as a mirror plane if we neglect multiple
scattering effects.\cite{AgSiH14} Solid lines indicate the binding
energies of the SS1 and SS2 states. As one can observe, intensity
of both states increases with $\Theta_{i}$. Nevertheless, the
intensity from SS1 increases slower than that from the SS2. In
fact, the SS1 increase is the 70\% of that of SS2, as calculated
after background subtraction. These results can be explained if we
consider that SS1 and SS2 are bulk-induced surface states. The
Ag(111) \textsl{sp}-gap is a Shockley inverted gap. In this case,
\textsl{sp}-surface states inside this gap are, basically,
\textsl{s}-like at the top and \textsl{p}-like at the bottom. If
we assume SS1 and SS2 to be \textsl{sp}-surface states, they are
expected to show a rather \textsl{p}-like behavior since they
appear quite close to the bottom of the gap. Therefore, the
observed increase of the SS1 and SS2 intensity with $\Theta_{i}$
can be attributed to their \textsl{p$_z$}-character. In any case,
if they were \textsl{only} \textsl{p}-like states, they should
show the same $\Theta_{i}$-dependence of the intensity.
Nevertheless, SS1 shows a smoother $\Theta_{i}$-dependence than
that of SS2 which can be attributed to a non-negligible
\textsl{s}-component. This contribution would be expected to be
more important for \textsl{sp}-surface states at the top of the
gap, which is coherent with our results.

The behavior of the two different SS1 and SS2 features as coming
from the \textsl{sp}-surface state seems to be established. The
question now, is to determine the origin of the splitting of the
Ag \textsl{sp}-surface state into the SS1 and SS2 features. On one
side, our results show that annealing processes make an
irreversible decrease of the emission from SS2, whereas that from
SS1 increases (Fig.~\ref{fig3}). On the other side, it has been
observed that Ag deposited onto H/Si(111)-(1x1) surfaces at RT
partially removes H from the metal-semiconductor interface and
this desorption process is enhanced at higher deposition
temperatures.\cite{AgSiH17,AgSiH18,AgSiH19} This behavior has been
also observed in Cu and Au films deposited onto these
substrates.\cite{Amina,AgSiH19} A similar H-desorption mechanism
has also been observed when the studied films RT-deposited were
annealed.\cite{AgSiH18,Nishiyama} These facts suggest that the SS2
state is related to the presence of H at the metal-semiconductor
interface, which somehow makes the bulk Ag(111) surface state
appear closer to the \textsl{sp}-band gap edge.

Several effects can cause a shift of the \textsl{sp}-surface
state. In the following, we will discuss the different
possibilities for H to induce such effect. Let us now discuss the
possibility of the existence of different Ag relaxed regions in
films deposited onto H/Si(111) surfaces. It has been observed that
the binding energy of the \textsl{sp}-surface state can differ
from one substrate to other, depending on strain.\cite{karsten}
Nevertheless, this seems not to be the case of films as-deposited
onto H/Si(111)-(1x1), where only two splitted surface states are
observed. If this splitting was induced by strain, the Ag films
would be mostly composed by two differently strained domains,
which would imply that different points of the Ag BZ are
simultaneously probed by ARPES. This fact would make that the
measured bulk Ag \textsl{4d} features appeared anomalously
broadened, which is not observed.

The study of the nature of surface states can be analyzed in the
workframe of the phase-accumulation model,\cite{phaseaccmod} based
on the idea that surface states can be considered as electrons
trapped between the surface gap and the surface
barrier.\cite{Echenique} In this model, differences of the binding
energy of surface states are expected via the surface-barrier and
crystal phases.

In principle, there are two possibilities for H to modify the
binding energy of the bulk Ag(111) surface state. The first one is
related to the fact that Ag deposition at RT removes H from the
interface, which seems to migrate to the metal
surface.\cite{AgSiH19} This H on the surface appears to be mostly
removed when films are annealed at high
temperatures.\cite{AgSiH19} Under these conditions, the presence
of H on the metallic surface could induce a shift of the surface
state to higher binding energies ($\Delta$SS) as far as H produced
a decrease of the work-function of the metal at the surface and,
consequently, a decrease of the surface-barrier
phase.\cite{phaseaccmod} In fact, these effects have been observed
after adsorption of submonolayer alkali-metals on Cu(111) and on
Ag(111) films deposited onto graphite and Si(111)-(7x7) at low
temperature.\cite{karsten,alkali} In these cases, the magnitude of
the observed $\Delta$SS was a $\sim$10-20\% of the work-function
decrease. Nevertheless, in contrast to the observed role of alkali
metals on metallic surfaces, it has been observed that the
presence of H at the Ag(111) and Ag(110) surfaces increases the
metal work-function at the surface by 0.2-0.3
eV,\cite{plummer1,plummer2} slightly shifting the Ag
\textsl{sp}-surface state to lower binding energies. These facts
suggest that the splitting of the \textsl{sp}-surface state can
not be attributed to the presence of H at the metal surface.

The second possibility for H to modify the binding energy of the
Ag(111) \textsl{sp}-surface state comes from the presence of
absorbed H at the metal-semiconductor interface. After Ag
deposition, and mostly after annealing processes, H can be
progressively desorbed from the interface. Nevertheless, large
H-covered areas of Si can stay underneath the Ag film, since, when
one initial Ag cluster is formed, Ag adatoms appear not to be
spontaneously substituted for H atoms around Ag adsorption
sites.\cite{Sakurai} This situation would induce two different
limiting conditions as far as the surface-state wave-function
could be influenced by surface conditions. It has been early
observed that Ag \textsl{sp}-surface state shifts to lower binding
energy as thicker Ag films are deposited onto 1ML-Au/Ag
substrates,\cite{decaySS} recovering a bulk-like situation in
$\sim$10ML-thick Ag films. This fact was explained by the presence
of the interlayer Au film, which downshifts the bulk-like Ag
surface state. It those systems, the decay length of the surface
state was determined to spread along several MLs (28
\AA).\cite{decaySS} In our case, a similar behavior may be
expected in Ag films deposited onto H/Si(111)-(1x1) surfaces,
where H appears to introduce a larger shift of the surface state
than that observed in the Ag/1ML-Au/Ag system. This effect could
be also understood in the frame of the phase-accumulation
model.\cite{phaseaccmod} The presence of H at the interface would
be expected to increase the reflectivity of the
\textsl{sp}-surface state, due to its long decay length compared
to the film thickness.\cite{decaySS} This fact would induce an
increase of the effective crystal phase and, therefore, the
\textsl{sp}-surface state would downshift. In order to estimate
the magnitude of this effect, an increase of the crystal phase of
round to 30-40\% would downshift the sp-surface state in $\sim$0.2
eV, which would reproduce quite well the behavior observed in our
data for Ag films deposited onto H/Si(111)-(1x1) substrates.

\subsection{Fermi surface}

In this section we analyze the electronic properties of the Ag
films beyond the metallization onset, centering this study on the
determination of the FS of these films. The image in
Fig.~\ref{fig7} shows the photoelectron angular distribution
measured at the E$_F$ with h$\upsilon $=32 eV in a 6 ML Ag film
deposited onto H/Si(111)-(1x1) and then annealed to 300 $^{o}$C
for 5 min. Photoelectron angular distribution measurements were
carried out by selecting x-axis corresponding to the $\left[
\overline{1}10\right] $ direction of k$_{\Vert }$. The image has
been scaled in such a way that it is linear in photoemission
intensity and in k$_{\Vert }$. The photoemission intensity is
maximum for the brightest feature and minimum for the darkest one.
In this image, well-defined features are labelled, indicating the
momentum distribution, as a function of k$_{\Vert }$$_y$ and
k$_{\Vert }$$_x$, of initial states lying at the E$_F$. The Ag SBZ
contours are also plotted on the image, following the extended
zone scheme.

In the center of the image it can be observed two different
features, labelled as SS1 and SS2. The SS1 feature appears as an
intense spot at the center of the image of $\sim$0.10 \AA$^{-1}$
in size, whereas the SS2 feature is a homogeneous ring-like
feature centered at $\overline{\Gamma}$-point with a radius of
0.23 \AA$^{-1}$. These features appear at the FS cut as a
consequence of the crossing of the E$_F$ of the SS1 and SS2
surface states, as these states disperse. The fact that these
states show such a circular symmetry in the k$_{\Vert }$-plane
indicates that these states are isotropic parabolic states, as
expected for \textsl{sp}-derived surface states in bulk metallic
systems.\cite{Hufner}

Besides to the surface-state-related features, additional ones
appear in the image, at higher k$_{\Vert }$ values, which have
been labelled as FS1, FS2, and FS2'. The FS1 feature appears as a
distorted ringlike inside the first SBZ, which corresponds to the
cut of the bulk Ag(111) FS in the first BZ.\cite{juan2} From this
feature, the k$_F$ values of the FS of these Ag films can be
extracted. In order to do this, one should take into account
experimental energetic-window effects, which move the E$_F$
band-crossing-point away from the maximum intensity of emitted
photoelectrons at the E$_F$ towards the unoccupied-states
side.\cite{AgSiH22,straub} In our experiments, the energetic
window was 0.2 eV with h$\upsilon $=32 eV. From this analysis, it
was obtained that this Ag FS cut shows a maximum and minimum k$_F$
values of k$_F$=1.29 and 1.16 \AA $^{-1}$, which occur along the
[$\overline{1}$10] and [$\overline{2}$11] (or
[$\overline{1}\overline{1}$2]) directions, respectively. These
values are in concordance with those obtained in metallic Ag films
prepared onto Si(111)-(7x7) surfaces.\cite{juan2} The resemblance
between the FS of both Ag films appears not only in the k$_F$
values as well as in the lifetime of \textsl{sp}-states at the
E$_F$.\cite{AgSiH22} Intrinsic photoemission line width of states
at the E$_F$ is dominated by photoelectron lifetime.\cite{smith}
In our experiments, the line width of the FS1 feature, where it
can be isolated from other contributions, is of 0.28$\pm$0.04 \AA
$^{-1}$ in average, which implies that the inverse lifetime of
photoelectrons is of 4.0$\pm$0.5 eV,\cite{AgSiH22} coherently with
that expected in the bulk material for photoelectrons excited with
h$\upsilon $=32 eV.\cite{AgSiH22,Goldmann}

The rest of features observed in Fig.~\ref{fig7} (FS2 and FS2')
appear outside of the first SBZ, as a semi-circular feature (FS2)
below a wide spot-like feature (FS2`). The presence of these
features is expected from neighboring BZs, which would be measured
with h$\upsilon $=32 eV.\cite{juan2} However, the most remarkable
electronic information to be extracted from the FS of these
metallic films is that it shows a sixfold symmetry, instead of the
typical threefold symmetry found in the three-dimensional Ag
single crystal. The loss of symmetry of the FS of these films can
be related to the fact that Ag films deposited onto
H/Si(111)-(1x1) at RT are composed by two domains rotated 60$^o$.

In order to illustrate this point, we have analyzed the influence
of the two-domains character of the films on the resulting Ag(111)
FS. Figure~\ref{fig8} shows (on the left) the band dispersion of
the \textsl{sp}-band of Ag(111) single crystal along the
high-symmetry directions of the SBZ: [$\overline{1}$10]
(k$_{\|x}$), [$\overline{2}$11] (k$_{\|y}$), and
[$\overline{1}\overline{1}$2], as would be obtained by ARPES with
h$\upsilon $=32 eV.\cite{juan2} In this plot, the crossing-points
of the \textsl{sp}-band with the E$_F$ are indicated, which are
labelled as e$_1$-e$_5$. Each one of these points defines one
point of the Ag(111) FS cut (middle plot) and, extending this to
the whole BZ, a two-dimensional image of the Ag FS can be
obtained. This FS cut of Ag single crystal shows the intrinsic
threefold symmetry of the Ag BZ. The obtention of such a FS by
ARPES assumes that each direction of the hexagonal Ag SBZ is
scanned along one experimental azimuth (upper plot). Nevertheless,
this seems not to be the case of the Ag films prepared onto
H/Si(111)-(1x1) at RT. A film composed of two domains rotated
60$^o$ would artificially produce, by ARPES, a 60$^o$-overlapping
of the band structure, of the SBZ, and, therefore, of the FS
(right plots in Fig.~\ref{fig8}). Since one experimental azimuth
simultaneously scans two 60$^o$-differing directions of the Ag
SBZ, belonging each one to two different domains.

The fact that Ag films deposited onto H/Si(111)-(1x1) at RT and
then annealed to 300 $^o$C for 5 min are composed by two domains
rotated 60$^o$ has been also observed in Ag films of similar
thicknesses deposited onto Si(111)-(7x7).\cite{juan2}
Nevertheless, in contrast to the results observed in films
deposited onto H/Si(111)-(1x1) at RT, it seems that
high-temperature Ag deposition favors a mono-domain
growth.\cite{AgSiH17} These facts suggest that there exists an
activation energy for the Ag nucleation beyond that mono-domain Ag
growth is favored. In addition to this, the fact that annealing
process carried out after deposition appears not to favor this
structural transition seems to confirm the fact observed that Ag
deposition rate may also play a role in the obtention of
mono-domain films.\cite{Nishiyama} Nevertheless, much work should
be done to elucidate this point. In any case, these results are
promising and suggest that a good knowledge and control of the
growth conditions of Ag films onto H/Si(111)-(1x1) substrates may
improve the conductivity and transport properties of these films,
since the influence of electronic barriers and hopping processes
would be reduced.

\section{Summary}

Silver films of different thickness (0-8 MLs) were deposited at RT
onto H-passivated Si(111) surfaces. Their electronic properties
have been analyzed by ARPES, probing $\overline{K}$ and
$\overline{\Gamma}$-points of the Si(111) SBZ. The valence-band
spectra of submonolayer Ag films appeared to follow the evolution
expected for an abrupt and nonreactive metal-semiconductor
interface, with no-trace observed from Ag-induced surface states
at these points of the SBZ. Bulk-like metallization process was
followed by analyzing the evolution of the DOS at the E$_F$ at
Si(111) $\overline{K}$-point, which coincides, at h$\upsilon $=32
eV, with the crossing-point of the bulk Ag \textsl{sp}-band with
the E$_F$. At initial Ag growth-stage, films appeared to be
semiconducting and the onset of metallization was established at a
coverage of $\sim$0.6 MLs. The bulk-like metallic behavior was
completely defined in Ag films of 2-3 MLs and the small line width
of the Ag \textsl{4d}-derived features indicates that high-quality
(111)-oriented films can be prepared onto H-passivated Si
surfaces.

Two well-defined Ag-derived features appeared in the valence-band
spectra measured at $\overline{\Gamma}$-point, in films thicker
than 2 MLs. These features, labelled as SS1 and SS2, were assigned
to occupied states with binding energies of 0.1 and 0.35 eV,
respectively. These states were both identified, by means of their
perpendicular and parallel dispersion, as two surface states
appearing at the \textsl{sp}-gap of the Ag SBZ and related to the
\textsl{sp}-surface state of bulk Ag(111). As expected for such
surface state, the wave function of both SS1 and SS2 states shows
a pronounced \textsl{p}-like behavior, as revealed by
photoemission measurements tuning the components of the
polarization vector of light. The existence of two different
surface states with the same origin was attributed to the
non-homogeneous presence of H at the interface, due to the fact
that Ag deposition and annealing process partially desorb H from
the interface and, therefore, a downshift of the
\textsl{sp}-surface state is produced in regions where H is still
present.

The electronic properties of metallic Ag films deposited at RT
onto H-passivated substrates were analyzed by measurements of
photoelectron angular distribution at the E$_F$ with h$\upsilon
$=32 eV. With this technique, a cross-sectional cut of the FS of
these films has been obtained with this h$\upsilon $. The k$_F$
values of this FS cut obtained in different symmetry directions
appeared to be in good agreement with those expected for bulk-like
Ag(111) single crystal. A similar agreement was also found for the
photoelectron lifetime of states at the E$_F$. In spite of the
fact that the FS cut resembles a two-dimensional cut of the
three-dimensional bulk-like Ag FS, it reflects a sixfold symmetry
rather than the threefold symmetry expected for Ag single crystal.
This indicates that these Ag films are composed by two domains
rotated 60$^o$.

\acknowledgments

This work was financed by DGICYT (Spain) (Grant No. PB-97-1199)
and the Large Scale Facilities program of the EU to LURE.
Financial support from the Comunidad Aut\'{o}noma de Madrid
(Project No. 07N/0042/98) is also acknowledged. A.A. and J.F.S.-R.
acknowledge financial support from the Ministerio de Educaci\'{o}n
y Cultura of Spain.


\begin{figure}
\caption{EDCs measured with h$\upsilon$=32 eV in Ag films
deposited at RT onto H/Si(111)-(1x1) surfaces, at: (a) Si(111)
$\overline{K}$-point and (b) $\overline{\Gamma}$-point. The
corresponding Ag coverage is indicated on each curve. The
different labelled features are those corresponding to the
substrate. The insets show in detail the evolution with coverage
of the DOS at the E$_F$.} \label{fig1}
\end{figure}

\begin{figure}
\caption{EDCs measured with h$\upsilon$=32 eV in Ag films
deposited at RT onto H/Si(111)-(1x1) surfaces, at: (a) Si(111)
$\overline{K}$-point and (b) $\overline{\Gamma}$-point. The
corresponding Ag coverage is indicated on each curve. Labelled
features are those of Fig.~\ref{fig1}. The insets show in detail
the evolution with coverage of the DOS at the E$_F$. Two features,
labelled as SS1 and SS2, have been identified at
$\overline{\Gamma}$-point close to the E$_F$.} \label{fig2}
\end{figure}

\begin{figure}
\caption{Normal-emission valence-band spectra measured with
h$\upsilon $=32 eV in different Ag films deposited at RT onto
H/Si(111)-(1x1). Curves \textsl{a}-\textsl{c} correspond,
respectively, to 4, 5, and 6 ML-thick as-deposited Ag films. Curve
\textsl{d} corresponds to the latter 6 ML-thick Ag film annealed
to 300 $^{o}$C for 5 min. Curve \textsl{e} corresponds to the
latter one annealed again to 300 $^{o}$C for 15 min and then 1
ML-thick Ag film was deposited. Curve \textsl{f} corresponds to a
further Ag deposition of 1 ML and then annealed to 300 $^{o}$C for
5 min. The binding energies of the SS1 and SS2 states are
indicated by solid lines in these spectra.} \label{fig3}
\end{figure}

\begin{figure}
\caption{Valence-band spectra measured by ARPES along the
[$\overline{1}$10] direction with h$\upsilon$=32 eV in a 7
ML-thick Ag film deposited at RT onto H/Si(111)-(1x1) and annealed
to 300 $^{o}$C for 20 min. Small solid bars indicate the binding
energies of the SS1 and SS2 states as they disperse. The inset
shows the band diagram as extracted from the dispersion of these
features. Solid lines are parabolic fitting curves of the obtained
dispersions.} \label{fig4}
\end{figure}

\begin{figure}
\caption{Normal-emission valence-band spectra measured in a 6
ML-thick Ag film deposited at RT onto H/Si(111)-(1x1) and then
annealed to 300 $^{o}$C for 5 min, as a function of the
h$\upsilon$. Solid lines indicate the binding energies of the SS1
and SS2 states in these spectra.} \label{fig5}
\end{figure}

\begin{figure}
\caption{Normal-emission valence-band spectra measured with
h$\upsilon$=50 eV in a 6 ML-thick Ag film deposited onto
H/Si(111)-(1x1) and then annealed to 300 $^{o}$C for 5 min, as a
function of the incident angle of light $\Theta_{i}$. Solid lines
indicate the binding energies of the SS1 and SS2 states.}
\label{fig6}
\end{figure}

\begin{figure}
\caption{(color) Photoelectron angular distribution measured at
the E$_F$ with h$\upsilon $=32 eV in a 6 ML Ag film deposited onto
H/Si(111)-(1x1) and then annealed to 300 $^{o}$C for 5 min. The
different features observed have been labelled in the image. The
hexagonal Ag(111) SBZ is plotted onto the image.} \label{fig7}
\end{figure}

\begin{figure}
\caption{On the left. Dispersion of the bulk Ag(111)
\textsl{sp}-band along the [$\overline{1}$10] (k$_{\|x}$),
[$\overline{2}$11] (k$_{\|y}$), and [$\overline{1}\overline{1}$2]
that would be obtained by ARPES with h$\upsilon $=32 eV. The
crossing-points of \textsl{sp}-band with the E$_F$ are indicated
and labelled as e$_1$-e$_5$. These points are also indicated on
the corresponding Ag(111) FS cut (middle plot). The hexagonal SBZ
is also plotted and high-symmetry directions are indicated (upper
plot). On the right. (cross, dots) Band dispersion that would be
obtained in a situation corresponding to Ag films with two domains
rotated 60$^o$. The resulting FS cut and SBZ are also plotted.}
\label{fig8}
\end{figure}

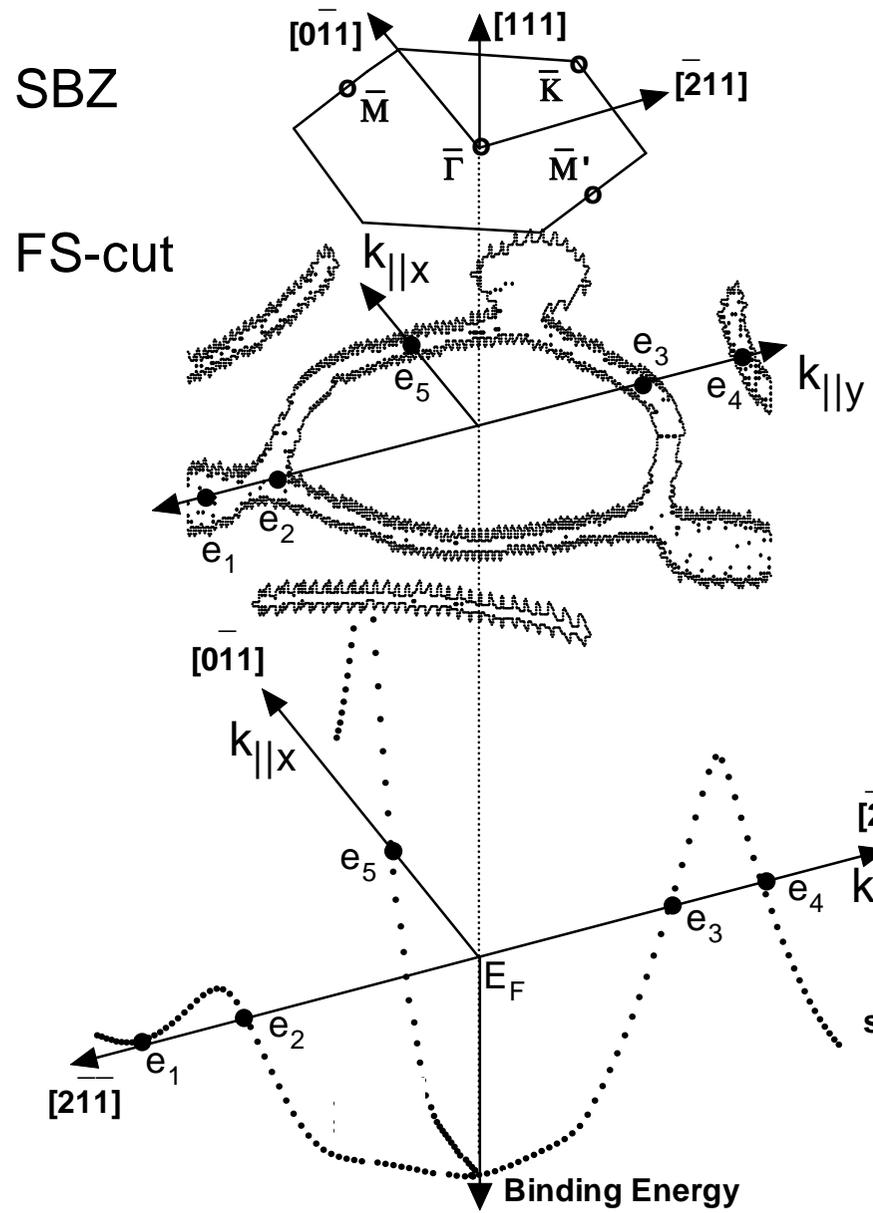
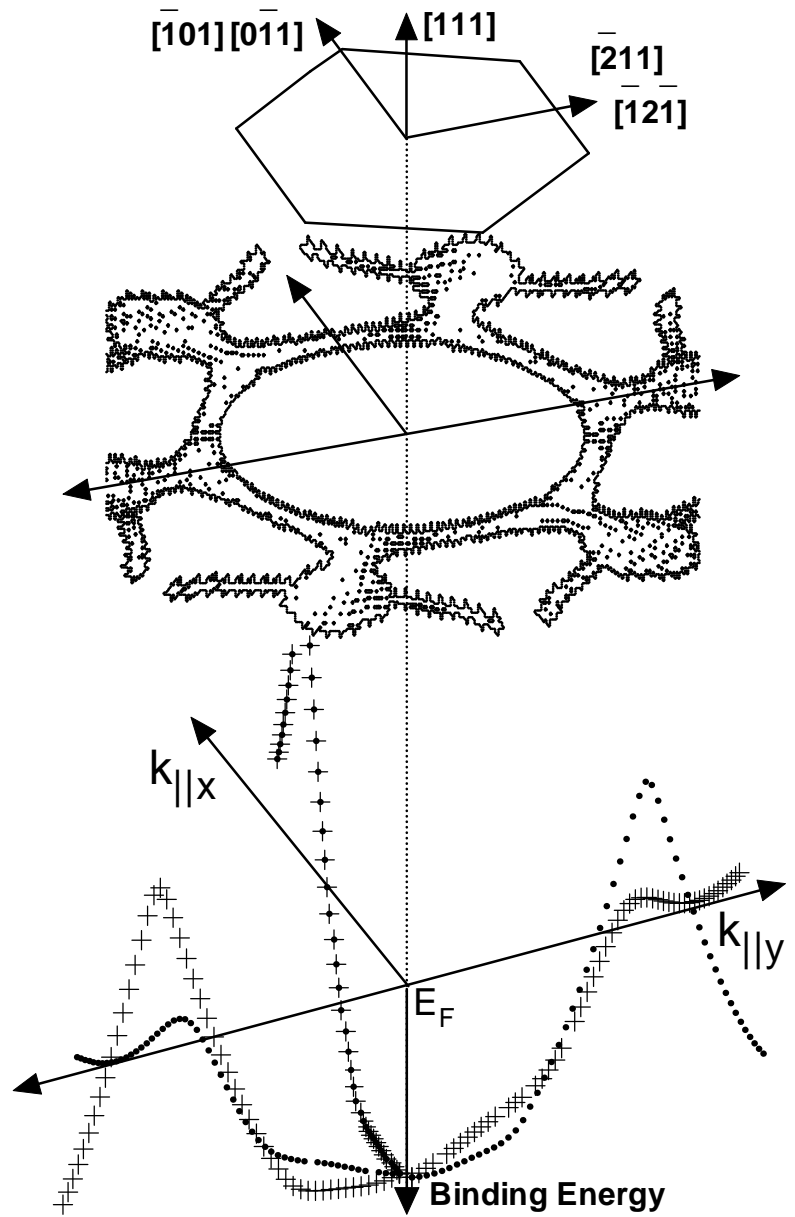
\end{document}